\documentclass[a4paper,11pt,oneside,final]{article}

\usepackage{amsmath,amssymb,bm,mathptmx,graphicx}

\usepackage[small,bf]{caption}

\usepackage{hyperref}
\hypersetup{pdftitle={Exact field-driven interface dynamics in the two-dimensional stochastic Ising model with helicoidal boundary conditions}, pdfauthor={J. R. G. Mendonça}, pdfsubject={Statistical mechanics, stochastic Ising model, exact solution}, pdfkeywords={Stochastic Ising model, interface dynamics, RSOS model, zero range process, exclusion process, XXZ model, Bethe ansatz, KPZ universality class}, unicode, extension=pdf, pdfnewwindow, pdftoolbar, pdfmenubar, pdffitwindow=false, pdfstartview=FitH, colorlinks, linkcolor=blue, citecolor=blue, urlcolor=blue}
\def\eps{\epsilon}

\def\leq{\leqslant}
\def\geq{\geqslant}
\def\I{{\rm i}}
\def\1{{\bf 1}}

\def\abs#1{\lvert{#1}\lvert}
\def\bra#1{\lvert{#1}\rangle}

\renewcommand{\thefootnote}{\fnsymbol{footnote}}
\def\starnote#1{\begingroup%
   \def\thefootnote{$\bigstar$}\footnote[0]{#1}%
\endgroup}

\begin{document}

\pagestyle{empty}

\begin{titlepage}

\vspace*{\fill}

\begin{center}

{\Large \bf Exact field-driven interface dynamics in the two-dimensional \\ stochastic Ising model with helicoidal boundary conditions}\starnote{This paper is dedicated to Professor S\'{\i}lvio R. A. Salinas (IF/USP) on the occasion of his 70th birthday.}

\vspace{6ex}

{\large {\bf J. Ricardo G. Mendon\c{c}a}$^{a,b,}$\footnote{Email: {\tt \href{mailto:jricardo@infis.ufu.br}{\nolinkurl{jricardo@infis.ufu.br}}}}}

\vspace{1ex}

$^{a}${\it {Instituto de F\'{\i}sica, Universidade Federal de Uberl\^{a}ndia \\ Caixa Postal 593, 38400-902 Uberl\^{a}ndia, MG, Brazil}}

\vspace{1ex}

$^{b}${\it {Instituto de F\'{\i}sica, Universidade de S\~{a}o Paulo \\ Caixa Postal 66318, 05314-970 S\~{a}o Paulo, SP, Brazil}}

\vspace{6ex}

{\large \bf Abstract \\}

\vspace{2ex}

\parbox{120mm}{
We investigate the interface dynamics of the two-dimensional stochastic Ising model in an external field under helicoidal boundary conditions. At sufficiently low temperatures and fields, the dynamics of the interface is described by an exactly solvable high-spin asymmetric quantum Hamiltonian that is the infinitesimal generator of the zero range process. Generally, the critical dynamics of the interface fluctuations is in the Kardar-Parisi-Zhang universality class of critical behavior. We remark that a whole family of RSOS interface models similar to the Ising interface model investigated here can be described by exactly solvable restricted high-spin quantum XXZ-type Hamiltonians.

\vspace{2ex}

{\noindent}{\bf Keywords}: Stochastic Ising model~$\cdot$ RSOS growth model~$\cdot$ zero range process~$\cdot$ exclusion process~$\cdot$ XXZ quantum chain~$\cdot$ Bethe ansatz~$\cdot$ KPZ universality class

\vspace{2ex}

{\noindent}{\bf PACS 2010:} 05.20.--y~$\cdot$ 05.40.--a~$\cdot$ 64.60.Cn~$\cdot$ 64.60.Ht

\vspace{2ex}

{\noindent}{\bf Journal ref.:} \href{http://dx.doi.org/10.1016/j.physa.2012.07.069}{{\it Physica A\/} {\bf 391}, 6463--6469 (2012)}.
}

\end{center}

\vspace*{\fill}

\end{titlepage}


\pagestyle{plain}

\section{\label{intro}Introduction}

Mappings from two-dimensional (2D) Ising model interface configurations to diffusion processes are known at least since the work of Rost \cite{rost}, and have been explored many times since then \cite{marchand,kandel,nienhuis,ferreira,majumdar,barma,%
spohn,paczuski,abraham,strobel}. In some cases \cite{marchand,kandel,nienhuis}, the interface dynamics of the model at zero temperature in the absence of an external field was mapped into the one-dimensional symmetric simple exclusion process, with the main result being the solution of a first-passage time problem showing that the mean lifetime of a shrinking domain is proportional to its initial area, providing a microscopic derivation for this well known experimental fact. It was also recognized that the resulting exclusion process can be recast as a probabilistic cellular automaton with a transition matrix equivalent to the transfer matrix (in a diagonal direction) of the symmetric six-vertex model in one of its critical lines \cite{kandel,nienhuis}.

The relationship between interfaces, exclusion processes, and vertex models was explored further \cite{dhar,krug,gwa,kim}, and it was realized that a Heisenberg Hamiltonian with pure imaginary Dzya\-lo\-shins\-ky-Moriya interaction that commutes with the transfer matrix of a six-vertex model describes the single-step surface growth model \cite{meakin,plischke}, as well as a discrete-velocity version of the noisy Burgers equation, which in turn is equivalent to the Kardar-Parisi-Zhang equation \cite{kpz}.
In the interacting particle system scenario, the above mentioned Hamiltonian is but the infinitesimal generator of the asymmetric simple exclusion process \cite{dhar,gwa}. Conversely, a host of results concerning symmetric and asymmetric simple exclusion processes has been translated into the Ising interface problem and, in particular, the investigation of the motion of tagged particles, as first introduced in the study of the hydrodynamic behavior of exclusion-type processes \cite{demasi,pablo}, has provided a partial explanation for the relationship between the characteristics of different asymptotic growth regimes in some $(1+1)$-dimensional stochastic growth models \cite{majumdar,barma,paczuski}.

In this article we show that under suitable generalized, but otherwise quite natural periodic boundary conditions the dynamics of an interface in the 2D stochastic Ising model in the presence of an external driving field can be mapped via a particle-height transformation into the dynamics of hopping particles without exclusion known as the zero range process \cite{spitzer,zero,andjel,bjp,euro,wadati,povolot}. The infinitesimal generator of the zero range process is equivalent to a high-spin, in general asymmetric quantum Hamiltonian that is exactly solvable by the Bethe ansatz \cite{vertex,mixture,bariev}.  We argue that the critical behavior of a generalized particle-height model must be on the Kardar-Parisi-Zhang universality class of critical behavior, since this is the critical behavior of the corresponding generalized quantum chains. This may have implications in the study of related models such as the dynamics of $k$-mers and other Ising-type lattice configurations.

The article is organized as follows. In Section~\ref{model} we introduce the 2D Ising model in an external field and the single-spin flip rates in terms of which the dynamics of the Ising contours will be analyzed, and in Section~\ref{mapping} we show that it can be described by diffusing particles without exclusion and exhibit the infinitesimal generator of the process. Section~\ref{solve} contains a brief exposition of the exact solution of the zero range process by the Bethe ansatz and a discussion on its dynamical critical exponent. In Section~\ref{xxs}, we show that a whole class of interface models similar to the Ising interface model can be described by exactly solvable, generalized restricted XXZ-type Hamiltonians with many modeling possibilities. Finally, in Section~\ref{SUMM} we summarize our results and indicate some directions for further investigation.


\section{The 2D stochastic Ising model in a field}
\label{model}

The 2D Ising model in an external field is described by
the Hamiltonian
\begin{equation}
\label{HN}
H({\bf S}) = -J\sum_{\langle{{\bf r},{\bf r}'}\rangle}
S_{{\bf r}}S_{{\bf r}'}-B\sum_{{\bf r}}S_{{\bf r}},
\end{equation}
where ${\bf S} = \{S_{{\bf r}}: {{\bf r}} \in \Lambda_{L}^{N}\}$ with $S_{{\bf r}} \in \{-1,+1\}$ are Ising spins, $\Lambda_{L}^{N} \subset {\mathbb Z}^{2}$ is a member of a family of semi-infinite lattices of $\abs{\Lambda_{L}^{N}}=L \times \infty$ sites, and $\langle{{\bf r},{\bf r}'}\rangle$ denotes pairs of nearest neighbor sites on $\Lambda_{L}^{N}$. The integer index $N$ in $\Lambda_{L}^{N}$ refers to the boundary conditions, that are free in the infinite direction and helicoidal with pitch $N$ in the finite direction, i.e., ${\bf r}+ L{\bf x}+N{\bf y} \equiv {\bf r}$ for all ${{\bf r}} \in \Lambda_{L}^{N}$. When $N=0$ we recover the usual periodic boundary condition, which is however uninteresting for our purposes, as we will see later. In the above Hamiltonian we take $J > 0$, making the model ferromagnetic, and for definiteness we take $B \geq 0$.

We introduce a dynamics on the Ising spins through the master equation
\begin{equation}
\label{MASTER}
\frac{{\rm d}}{{\rm d}t}P({\bf S},t) = \sum_{\tilde{{\bf S}} \in \Omega(\Lambda_{L}^{N})}
\bigg[ W(\tilde{{\bf S}} \to {\bf S}) P(\tilde{{\bf S}},t)-
 W({\bf S} \to \tilde{{\bf S}}) P({\bf S},t) \bigg]
\end{equation}
for the probability $P({\bf S},t)$ of observing the configuration ${\bf S} \in \Omega(\Lambda_{L}^{N}) = \{-1,+1\}^{\Lambda_{L}^{N}}$ at instant $t$, where $W({\bf S} \to \tilde{{\bf S}})$ is the rate at which configuration $\tilde{{\bf S}}$ is reached from configuration ${\bf S}$ per unit time. The rates $W({\bf S} \to \tilde{{\bf S}})$ should be translation invariant and verify the condition of detailed balance $W(\tilde{{\bf S}} \to {\bf S}) P(\tilde{{\bf S}}) = W({\bf S} \to \tilde{{\bf S}}) P({\bf S})$, with $P({\bf S}) \propto \exp[-\beta H({\bf S})]$ the Gibbs equilibrium probability distribution and where for the sake of notational economy we omitted the dependence of $P({\bf S})$ and $W({\bf S} \to \tilde{{\bf S}})$ on $J$, $B$, and the inverse temperature $\beta = 1/k_{\rm B}T$. In this work we consider heat-bath single-spin flip transition rates given by
\begin{equation}
\label{RATES}
W({\bf S} \to \tilde{{\bf S}}) \equiv W(S_{{\bf r}} \to \tilde{S}_{{\bf r}}) = \frac{1}{Z_{{\bf r}}} \exp[-\beta H(S_{{\bf r}})],
\end{equation}
with
\begin{equation}
H(S_{{\bf r}}) = -J\sum_{\langle {\bf r}':{\bf r} \rangle} 
S_{{\bf r}}S_{{\bf r}'}-BS_{{\bf r}} \quad {\rm and} \quad
Z_{{\bf r}} = \sum_{S_{{\bf r}} = \pm 1} \exp[-\beta H(S_{{\bf r}})],
\end{equation}
where $\langle{\bf r}':{\bf r}\rangle = \{{\bf r}' \in \Lambda_{L}^{N} : \abs{{\bf r}'-{\bf r}}=1\}$.

\begin{figure}
\centering
\includegraphics[viewport=69 135 526 650, clip, scale=0.42, angle=-90]{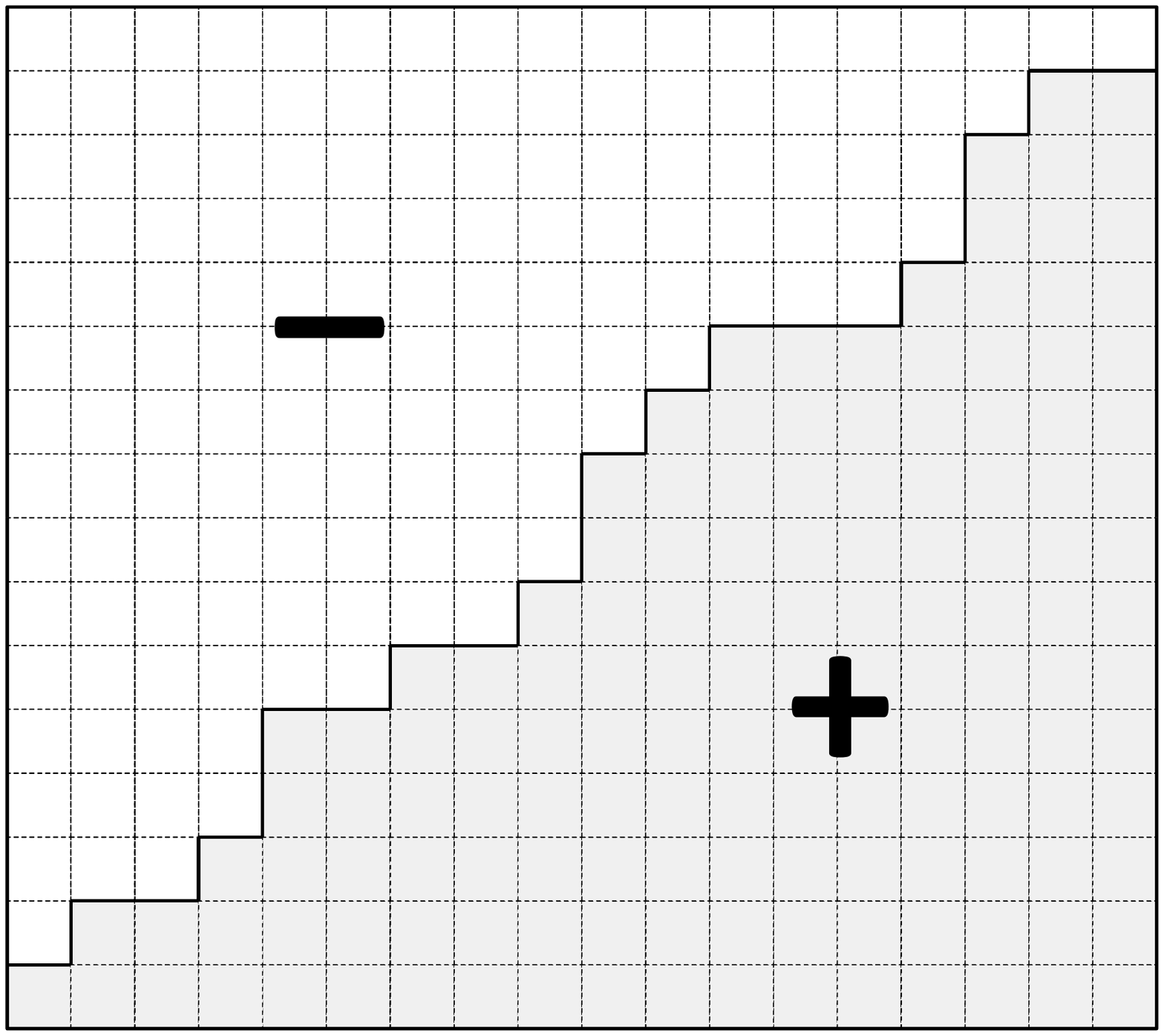}
\caption{Staircase-like 2D Ising interface separating the ``$+$'' and the ``$-$'' phases. In this figure, $L=18$ and $N=15$. When $B>0$ ($B<0$), the ``$+$'' (``$-$'') phase invades the ``$-$'' (``$+$'') phase, while for $B=0$ the interface only fluctuates about its initial shape. When $\beta \to \infty$ and $\abs{B} < 2J$, the number of bonds $N$ is conserved.}
\label{fig:interface}
\end{figure}

Let $w(S_{{\bf r}}) = \frac{1}{2} \sum_{\langle{{\bf r}':{\bf r}}\rangle} \abs{S_{{\bf r}'}-S_{{\bf r}}}$ be the number of spins neighboring $S_{{\bf r}}$ that have the sign opposite to it. In terms of this quantity, the single-spin flip rates read
\begin{equation}
\label{WS}
W(S_{{\bf r}} \to \tilde{S}_{{\bf r}}) =
\frac{1}{1+\exp[4\beta J(2-w(S_{{\bf r}})) + 2\beta BS_{{\bf r}}]}.
\end{equation}
At sufficiently low temperatures, as long as $B < 2J$ spins with $w(S_{{\bf r}})=0,1$ will hardly flip, because their transition rates become exponentially small when compared with the other rates, of the order of $\exp[-2\beta(2J-B)]$ at maximum. Processes with $w(S_{{\bf r}})=3,4$ correspond to fast processes, since at sufficiently low temperatures and again in the range $B<2J$ their rates become close to unity, $W(S_{\bf r}\to\tilde{S}_{{\bf r}}) \geq 1-\exp[-2\beta(2J-B)]$. In the low temperature limit and in the range $B < 2J$, thus, the heat-bath single-spin flip rates (\ref{WS}) define a process in which only spins with $w(S_{{\bf r}}) \geq 2$ have an appreciable flipping rate, and henceforth we ignore the flipping of spins with $w(S_{{\bf r}}) < 2$.\footnote{In realistic pseudo-two-dimensional $S = \frac{1}{2}$ Ising-like materials, e.g.\ in the antiferromagnetic compounds K$_2$XF$_4$ with X = Mn, Fe, Co, or Ni, $J/k_{\rm B} \sim$ 1--100\,K, such that $\beta J \gg 1$ implies $T < 1$\,K \cite{experimental}. The values for which $B < 2J$ thus lie in the range $B \lesssim 1.5$\,T, of the order of half the magnetic field strength of a typical medical MRI system.} Spins with $w(S_{{\bf r}})=3,4$, in turn, can be avoided by choosing initial configurations in which the ``$+$'' phase is separated from the ``$-$'' phase by a single-valued, non-self-intersecting staircase-like interface as in Figure \ref{fig:interface}. With initial configurations of this type and within the low temperatures and fields regime, we are left with a process in which only spins with $w(S_{{\bf r}})=2$ flip. Since spins with $w(S_{\bf r})=2$ lie at the interface, the above-defined spin flip dynamics actually defines an interface dynamics. In the next section we map this dynamics into an interacting particle system on the integers.


\section{Mapping to the zero range process}
\label{mapping}

Various possibilities exist to map the dynamics of 2D Ising interfaces into a system of interacting particles on the line. One possible map is obtained by associating with every vertical bond in the interface a particle and with every horizontal bond a hole \cite{kandel}. In this way we end up with a simple exclusion process in which particles hop in an augmented one-dimensional lattice. Another possibility was given in \cite{majumdar,barma}. In this case, one considers a set of $L$ particles on a one-dimensional lattice occupying the positions $x_{\ell}$, $1 \leq \ell \leq L$. If one associates the particle labels $\ell$ with a horizontal coordinate, and the particle positions $x_{\ell}$ with the heights of an interface, then one has a one-to-one map between the set of particles and an interface. With the additional constraint $x_{\ell+1}-x_{\ell} \geq 1$, the resulting model was called the particle-height model. The particle-height model thus establishes a map between the low-temperature dynamics of an Ising interface and the simple exclusion process, although the constraint on the particles positions seems a bit artificial in the interface scenario.

Our mapping of the Ising interface to a set of hopping particles on the integers is equivalent to the particle-height mapping with $x_{\ell+1}-x_{\ell} \geq 0$. Let $\Gamma_{L}^{N}$ be the set of all single-valued, non-self-intersecting staircase-like Ising interface configurations $\gamma_{L}^{N}$ of length $\abs{\gamma_{L}^{N}}=L+N$, $N \geq 0$, in the infinite strip $\Lambda_{L}^{N}$ of width $L$ with helicoidal boundary condition ${\bf r}+L{\bf x}+N{\bf y} \equiv {\bf r}$. The dynamics of the $\gamma_{L}^{N}$ interfaces under the action of the flipping rates $W(S_{{\bf r}})$ in the regime where $\beta \to \infty$ and $B < 2J$ preserves the length of the interfaces, i.e., given an initial configuration $\gamma_{L}^{N}(t=0) \in \Gamma_{L}^{N}$, all subsequent configurations $\gamma_{L}^{N}(t \geq 0) \in \Gamma_{L}^{N}$. The number $N$ of vertical bonds in the interface thus decomposes the state space of interface configurations in an infinite number of disjoint sectors, $\Gamma_{L} = \bigcup_{N \geq 0} \Gamma_{L}^{N}$.

The interface configurations $\gamma_{L}^{N}$ are single-valued functions with respect to the horizontal coordinate. Let $h_{\ell} \in {\mathbb Z}$, $1 \leq \ell \leq L$ denote the height of the Ising interface at site $\ell$. Then the heights differences $n_{\ell} = h_{\ell+1}-h_{\ell}$ are nonnegative due to the special form of $\gamma_{L}^{N}$, and their dynamics relates to the dynamics of the interface as follows. Let us consider the case $0 < B < 2J$, and that the ``$+$'' phase lies below the ``$-$'' phase, as in Figure~\ref{fig:interface}. In this way, every ``$-$'' (``$+$'') spin that flips contributes to the growth of the ``$+$'' (``$-$'') phase, increasing (decreasing) the height variable associated with its horizontal position by one unit. The single-spin flipping rates are given by
\begin{equation}
\label{rates}
- \to + : \: p=\frac{1}{1+e^{-2\beta B}}
\qquad {\rm and} \qquad 
+ \to - : \: q=\frac{1}{1+e^{2\beta B}} = 1-p.
\end{equation}
When $B=0$, we have a spin (time) reversal invariant system, and the interface does not move bodily. Otherwise, when $0 < B < 2J$ and $\beta \to \infty$, $p=1$ and $q=0$, and the surface can only grow. The general situation $0 < q < p < 1$ is obtained by taking $\beta \to \infty$ and $B \to 0$ with $\beta B={\rm constant}$. The heights differences moves corresponding to the flipping processes are $(n_{\ell-1},n_{\ell}) \to (n_{\ell-1}-1,n_{\ell}+1)$ when $h_{\ell} \to h_{\ell}+1$, and $(n_{\ell-1},n_{\ell}) \to (n_{\ell-1}+1,n_{\ell}-1)$ when $h_{\ell} \to h_{\ell}-1$. The maximum possible height difference in $\Lambda_{L}^{N}$ is $n_{\ell}=N$, in which case $h_{1}=h_{2}= \ldots =h_{\ell}=0$ and $h_{\ell+1}=h_{\ell+2}= \ldots =h_{L}=N$ for some $\ell$. More generally, we have $\sum_{\ell=1}^{L-1}n_{\ell}=h_{L}-h_{1}=N$. Now one appreciates the role of the helicoidal boundary conditions on $\Lambda_{L}^{N}$: the pitch $N$ gives the total number of interacting particles in the heights differences scenario. The boundary condition in the heights differences scenario is simply periodic.

The dynamics of the variables $n_{\ell}$ is but the zero range dynamics, in which particles hop on the lattice without exclusion \cite{spitzer,zero,andjel,bjp,euro,wadati,povolot}. Except for the (immaterial) absolute values of the heights, the dynamics of the $n_{\ell}$ variables contains all the information about the evolving Ising interface. The elementary processes for the $n_{\ell}$ variables are
\begin{equation}
\label{process}
(n_{\ell-1}+1,n_{\ell}) 
 \begin{array}{c} p \\ \rightleftharpoons \\ q \end{array} 
(n_{\ell-1},n_{\ell}+1), 
\quad 0 \leq n_{\ell-1},n_{\ell} \leq N-1, \quad 2 \leq \ell \leq L.
\end{equation}
As is well known \cite{adhr,schutz}, we may write the master equation for reaction-diffusion processes on the lattice as a Schr\"{o}dinger-like equation in Euclidean time, the infinitesimal generator of the Markov semigroup playing the role of the quantum Hamiltonian. In this scenario, the infinitesimal generator of the above zero range process is given by the $N^L \times N^L$ matrix operator
\begin{equation}
\label{ZERO}
H_{N}=\sum_{\ell=1}^{L}  \sum_{m+n=0}^{N-1}
\bigg[p\Big(E_{\ell}^{m+1,m+1}E_{\ell+1}^{n,n}-E_{\ell}^{m,m+1}E_{\ell+1}^{n+1,n}\Big)+
 q\Big(E_{\ell}^{m,m}E_{\ell+1}^{n+1,n+1}-E_{\ell}^{m+1,m}E_{\ell+1}^{n,n+1}\Big)\bigg],
\end{equation}
where $E_{\ell}^{m,n}=\1 \otimes \cdots \otimes \1 \otimes E^{m,n} \otimes \1 \otimes \cdots \otimes \1$, with $\1$ the $N \times N$ identity matrix and $(E^{m,n})_{i,j} = \delta_{i,m}\delta_{j,n}$ the $N \times N$ matrix with a single unit element in row $m$ and column $n$ occupying the $\ell$-th position in the direct product. The operator $H_{N}$ can be diagonalized by the coordinate Bethe ansatz and, indeed, it has been diagonalized by this and related methods many times in the literature \cite{vertex,mixture,bariev,lazo}. We will thus not reproduce a complete resolution of (\ref{ZERO}) here. Instead, we just outline the technique and quote the main results regarding the process defined by $H_{N}$ of interest to us.


\section{Bethe ansatz solution}
\label{solve}

\subsection{Bethe ansatz equations}

We are interested in the solutions of the eigenvalue equation
\begin{equation}
\label{HPSI}
H_{N}\bra{\Psi_{N}} = E_{N}\bra{\Psi_{N}},
\end{equation}
where $H_{N}$ is given in (\ref{ZERO}) and
\begin{equation}
\label{ansatz}
\bra{\Psi_{N}} = \sum_{x_{1}  \leq  x_{2}  \leq  \cdots  \leq  x_{N}}
\Phi(x_{1},x_{2},\ldots,x_{N})\bra{x_{1},x_{2},\ldots,x_{N}}
\end{equation}
is the eigenfunction written in the basis that specifies the positions of the $N$ particles in the system, with $\Phi(x_{1},x_{2},\ldots,x_{N})$ the coefficient for the configuration $\bra{x_{1},x_{2},\ldots,x_{N}}$. Notice that since there is no exclusion, particle positions can coincide.

If the positions of the particles obey $x_{j+1} > x_{j}$, $1 \leq j \leq N$,
the eigenvalue equation (\ref{HPSI}) is satisfied by the ansatz (\ref{ansatz})
with coefficients
\begin{equation}
\label{PHI}
\Phi(x_{1},x_{2},\ldots,x_{N})=\sum_{P}A_{P(1)P(2) \cdots P(N)}
\exp\bigg[\I\sum_{j=1}^{N}k_{P(j)}x_{j}\bigg]
\end{equation}
and eigenvalue
\begin{equation}
\label{EEN}
E_{N}(k_{1},k_{2},\ldots,k_{N})=\sum_{j=1}^{N}\eps(k_{j}),
\end{equation}
where $\eps(k) = 1-pe^{-\I k}-qe^{\I k}$ is the ``single particle energy,'' the first summation in (\ref{PHI}) is over all the $N!$ permutations $P$ of the indices $(1,2,\ldots,N)$ used to label the positions of the particles, and the ``wave numbers'' $k_{1},k_{2},\ldots,k_{N}$ are chosen so that $\bra{\Psi_{N}}$ satisfies (\ref{HPSI}). We see that when the particles are far apart, in the case being just not on the same site, they behave as if they were free, and the total ``energy'' of the system is the sum of the ``energies'' of single particles. When a pair of particles sit on the same site, $x_{j+1}=x_{j}$, we obtain from (\ref{HPSI})--(\ref{EEN}) that the amplitudes $A_{P(1)P(2) \cdots P(N)}$ should satisfy
\begin{equation}
\label{A1N}
\frac{A_{P(1)P(2) \cdots P(j)P(j+1) \cdots P(N)}}
 {A_{P(1)P(2) \cdots P(j+1)P(j) \cdots P(N)}}=
 -\frac{e^{\I k_{P(j+1)}}}{e^{\I k_{P(j)}}}e^{\I \Theta_{P(j)P(j+1)}},
\end{equation}
where the ``two-particle scattering phase'' $\Theta_{j\ell}$ is defined by
\begin{equation}
\label{PHASE}
e^{\I \Theta_{j\ell}}=\frac{p+qe^{\I (k_{j}+k_{\ell})}-e^{\I k_{j}}}
 {p+qe^{\I (k_{j}+k_{\ell})}-e^{\I k_{\ell}}}.
\end{equation}
The boundary condition $\Phi(x_{2},x_{3},\ldots,x_{1}+L)=\Phi(x_{1},x_{2},\ldots,x_{N})$ furnishes the additional relation
\begin{equation}
\label{PBC}
A_{P(1)P(2) \cdots P(N)}=e^{\I k_{P(1)}L}A_{P(2)P(3) \cdots P(1)}.
\end{equation}
Iterating relation (\ref{A1N}) $N$ times, (\ref{PBC}) gives us the Bethe ansatz equations for the ``wave numbers'' $k_{j}$ in the $N$-particle sector,
\begin{equation}
\label{BAEN}
e^{\I k_{j}L}=(-1)^{N-1}\prod_{\ell=1}^{N}
\bigg(\frac{e^{\I k_{\ell}}}{e^{\I k_{j}}}\bigg)e^{\I \Theta_{j\ell}}=
(-1)^{N-1}\prod_{\ell=1}^{N}\bigg(\frac{e^{\I k_{\ell}}}{e^{\I k_{j}}}\bigg)
\frac{p+qe^{\I (k_{j}+k_{\ell})}-e^{\I k_{j}}}{p+qe^{\I (k_{j}+k_{\ell})}-e^{\I k_{\ell}}},
\quad 1 \leq j \leq N.
\end{equation}
The solutions $k_{1}, k_{2}, \ldots, k_{N}$ of these equations give through (\ref{EEN}) the eigenvalues of~(\ref{ZERO}). Notice that since $H_N$ is in general nonhermitian, the $k_{j}$ are in general complex numbers.

The eigenfunctions (\ref{ansatz}) with the coefficients (\ref{PHI}) should also be eigenfunctions of the translation operator $T$ that shifts the positions of the particles to the left by one site, since $[H,T]=0$. The eigenvalues $e^{\I P}$ of $T$ are given by
\begin{equation}
\label{mom}
T\bra{\Psi_{N}} = e^{\I P}\bra{\Psi_{N}} =
\bigg(\prod_{j=1}^{N}e^{\I k_{j}}\bigg)\bra{\Psi_{N}},
\end{equation}
where we have defined the total momentum-like $P$ by
\begin{equation}
P=\sum_{j=1}^{N}k_{j} \ ({\rm mod}\ 2\pi) = \frac{2\pi\ell}{L},
\quad 0 \leq \ell \leq L-1.
\end{equation}
With $P$ defined above, equations (\ref{BAEN}) can be rewritten as
\begin{equation}
\label{asep}
e^{\I k_{j}(L+N)}=(-1)^{N-1}e^{\I \sum_{\ell=1}^{N}k_{\ell}}
\prod_{\ell=1}^{N}e^{\I \Theta_{j\ell}}= (-1)^{N-1}e^{\I P}\prod_{\ell=1}^{N}
\frac{p+qe^{\I (k_{j}+k_{\ell})}-e^{\I k_{j}}}{p+qe^{\I (k_{j}+k_{\ell})}-e^{\I k_{\ell}}},
\quad 1 \leq j \leq N.
\end{equation}
The learned reader will recognize in (\ref{asep}) the Bethe ansatz equations for the asymmetric simple exclusion process of $N$ particles in a lattice of $L+N$ sites with twisted boundary conditions, the angle of twist being given by the total momentum $P$ of the system. For stochastic processes, the relevant momentum sector is the $P=0$ sector, since the coefficients $\Phi(x_{1},x_{2},\ldots,x_{N})$ have to be all real and positive. In this sector, the correspondence between the zero range process and the asymmetric simple exclusion process is exact.

\subsection{The dynamical critical exponent}

Numerical simulations together with theoretical arguments and explicit
calculations indicate that the critical behavior of the interface is independent of the particular values of $p$ and $q$ as long as $p \neq q$ \cite{krug,gwa,kim,plischke}. It has then become usual to investigate the Bethe ansatz equations (\ref{BAEN}) with $p=1$, $q=0$, since this facilitates the analysis considerably. The more general cases $0 < q < p < 1$ were investigated in \cite{kim,mallick}. The $p=1$, $q=0$ case corresponds to a 2D Ising interface evolving in a finite field $B > 0$ but at zero temperature. However, at least for very low (but nonzero) temperatures, one has the same kind of critical behavior as observed at zero temperature \cite{devillard}. In the totally asymmetric simple exclusion process, another simplification of the Bethe ansatz equations comes with the choice of the half-filled sector $2N=L$. The analogous choice for the zero range process is to consider the sector with $N=L$, which corresponds in the interface scenario to an interface with average slope $\pi/4$.

The dynamical critical exponent $z$, that measures the degree of anisotropy between the spatial and temporal correlation lengths, can be determined from the asymptotic behavior of the gap $E_{N}^{(1)}(L)$ of $H_{N}$ through ${\rm Re}\{E_{N}^{(1)}(L)\} \sim L^{-z}$. For the asymmetric exclusion process with arbitrary $p \ne q$ and $\varrho = N/L$, the large $L$ asymptotic value of $E_{N}^{(1)}(L)$ is given by
\begin{equation}
\label{lll}
E_{N}^{(1)}(L) = -2C\abs{p-q}\sqrt{\varrho(1-\varrho)}L^{-3/2} \pm
2\pi\I \abs{(p-q)(1-2\varrho)}L^{-1},
\end{equation}
with an exact (numerically evaluated) $C = 6.509\,189\ldots$ \cite{mallick}. The dynamical critical exponent of the asymmetric exclusion process is then $z=3/2$, indicating that it belongs to the Kardar-Parisi-Zhang universality class of critical behavior \cite{kpz}.

For the zero range process, the available calculations of $E_{N}^{(1)}(L)$ are based on the analysis of slightly generalized models, with non-uniform hopping rates or in which particles can hop together \cite{wadati,povolot}. The totally asymmetric zero-range process with uniform rates in the $N=L$ sector was solved in \cite{mixture,bariev}, with the result that the first gap behaves like 
\begin{equation}
\label{zrp}
E_{N}^{(1)}(L) \sim a_{0}L^{-3/2} - \I \frac{\pi}{2} L^{-1},
\end{equation}
with an exact (numerically evaluated) $a_{0} = 2.301345\ldots$, first obtained in \cite{gwa}; see also \cite{kim}.\footnote{Notice that the constants $C$ in (\ref{lll}) and $a_{0}$ in (\ref{zrp}) are related by $C=2\sqrt{2}a_{0}$.} We thus see that, in either case, the gap behaves asymptotically as $E_{N}^{(1)}(L) \sim L^{-3/2}$, the dynamical critical exponent $z=3/2$, and both processes---the asymmetric simple exclusion process and the asymmetric zero-range process---belong to the Kardar-Parisi-Zhang universality class of critical behavior, and so does the driven interface dynamics of the 2D stochastic Ising model in the regime of low temperatures and fields.


\section{Generalized particle-height model and interface dynamics}
\label{xxs}

The particle-height model mentioned in Sec.~\ref{mapping} can be generalized
to processes in which the particle positions observe the constraint $x_{\ell+1}-x_{\ell} \geq s$, generating a process in which particles move only if the next particle is far apart by at least $s$ sites. Identifying a particle with the up spin state and a hole with the down spin state in the $\sigma^{z}$ basis, the time evolution of this restricted exclusion processes is governed by the infinitesimal generator \cite{mixture,bariev}
\begin{equation}
\label{hs}
H_{s}=-\sum_{\ell=1}^{L}
P_{s}\bigg[p\sigma_{\ell}^{-}\sigma_{\ell+1}^{+}+q\sigma_{\ell}^{+}\sigma_{\ell+1}^{-}
+\frac{1}{4}\big(\sigma_{\ell}^{z}\sigma_{\ell+1}^{z}-1\big)\bigg]P_{s},
\end{equation}
where $p$ and $q=1-p$ are the rates at which particles hop respectively to the right and to the left,
\begin{equation}
\label{proj-s}
P_{s} = \prod_{\ell=1}^{L} \bigg[\frac{1}{2}(1-\sigma_{\ell}^{z}) + 
\frac{1}{2}(1+\sigma_{\ell}^{z}) \prod_{j=1}^{s-1} \frac{1}{2}(1-\sigma_{\ell+j}^{z})\bigg]
\end{equation}
is the operator that projects out configurations in which particles are closer than by $s$ sites, and $\sigma^{\pm}=\frac{1}{2}(\sigma^{x} \pm \I\sigma^{y})$ and $\sigma^{z}$ are the usual Pauli spin-$\frac{1}{2}$ matrices.

The infinitesimal generator $H_{s}$ is exactly solvable by the Bethe ansatz for any choice of the integer $s$ \cite{mixture,bariev}. The $s=1$ case recovers the generator of the asymmetric simple exclusion process, while the generic $s > 1$ case describes, e.g., the driven diffusion of $s$-mers on the lattice \cite{kmers}. In fact, it has been shown that an arbitrary mixture of $s$-mers of different sizes, diffusing with the same rates but ruled by evolution operators $H_{s}$ with different $s$, can be integrated exactly, with the eigenspectrum depending only on the average size $\tilde{s}$ of the $s$-mers \cite{mixture,bariev}.

For the generalized particle-height model, either with a single type of particle or with an admixture of particles of different sizes, the simplifying filling fraction analogous to the condition $N=L$ in the zero-range process is given by $(1+\tilde{s})N=L$, where $\tilde{s}$ is the average size of the particles in the system, not necessarily a semipositive integer. Clearly, the larger the average particle size $\tilde{s}$, the smaller the average inteface slope $N/L = 1/(1+\tilde{s})$. In this case, the spectral gap of the process has been found to scale like \cite{mixture,bariev}
\begin{equation}
\label{mix-gap}
E_{N}^{(1)}(L) \sim a_{0}L^{-3/2} + 
\I \left(\frac{\tilde{s}-1}{\tilde{s}+1}\right) \frac{\pi}{2} L^{-1},
\end{equation}
with the same $a_{0} = 2.301345\ldots$ as before; compare with (\ref{zrp}). We can then predict that the interface dynamics obtained from the generalized particle-height model with $x_{\ell+1}-x_{\ell} \geq s$ also belongs to the Kardar-Parisi-Zhang universality class of critical behavior.

A most interesting thing would be to vizualize how the Ising interface configurations evolve in the case of a particle-height model that includes besides particles of positive sizes also particles with negative sizes, since in this case handles and loops could develop. More generally, it would be of interest to find physical applications of the operator (\ref{hs}) for negative values of $s$ or $\tilde{s}$, since they are all exactly solvable and display the same type of critical behavior.


\section{Summary and conclusions}
\label{SUMM}

We showed that it is possible to map the interface dynamics of the 2D stochastic Ising model in the regime of low temperatures and fields into an exactly solvable interacting particle system of hopping particles without exclusion. The infinitesimal generator of the process is exactly solvable by the Bethe ansatz, with an spectral gap in the asymmetric case scaling like $L^{-3/2}$ with the system size. The 2D Ising interface in the presence of a driving field then grows and fluctuates according to the Kardar-Parisi-Zhang universality class of critical behavior.  We remark that most studies (e.g., in the realm of nucleation dynamics) of 2D Ising interfaces are carried out in zero temperature, and several results were obtained in the absence of external fields. In our study, following \cite{ferreira,devillard}, we allow for finite temperatures and fields, as long as the conditions stated in section \ref{model} are met.  When $\beta = \infty$ ($T = 0^{+}$) or $B=0$, the process becomes symmetric and the Bethe ansatz analysis reduces to a simple spin-wave analysis. In this case the gap becomes $E_{N}^{(1)}(L) = 2\sin^{2}(\pi/L)$ (independent of $N$ as long as $N < L$), with asymptotic behavior $E_{N}^{(1)}(L \gg 1) = 2\pi^{2}L^{-2}$, and the interface only fluctuates, without moving or growing, according to the $z=2$ Edwards-Wilkinson universality class of critical behavior \cite{edwil}.

It would be desirable to explore the mapping of the Ising interface to the zero range process to investigate step-step correlation functions by tagged-particle methods within the context of exact Bethe solutions \cite{majumdar,barma,paczuski,demasi,pablo,andjel,aaf}, as well as the dynamics of special configurations like semi-infinite strips of the minority phase (``Ising fingers'') \cite{krapivsky} using some of the ideas exposed here. We hope to return to these subjects soon.


\section*{Acknowledgments}

This paper is dedicated to Professor S\'{\i}lvio Roberto de Azevedo Salinas (IF/USP) on the occasion of his 70th birthday on Oct.~25th, 2012. Through his personal vigor and commitment, he and his numerous graduate students created a productive school on statistical physics of model systems that helped to foster the scientific thinking and education in Brazil.

The author thanks Professor Francisco C. Alcaraz (IFSC/USP) for having called his attention to (\ref{hs}) and for helpful conversations. This work was partially supported by CNPq, Brazil, under grant PDS 151999/2010-4.


\end{document}